ATI 2015 - 70th Conference of the ATI Engineering Association

# LES of the Sandia Flame D Using an FPV Combustion Model

M. Di Renzo[a], A. Coclite[a], M. D. de Tullio[a], P. De Palma[a],*  and G. Pascazio[a]

[a]*Department of Mechanic, Mathematic and Management, CEMeC, Polytechnic of Bari, via Re David 200, Bari, Italy*

**Abstract**

The simulation of turbulent combustion phenomena is still an open problem in modern fluid dynamics. Considering the economical importance of hydrocarbon combustion in energy production processes, it is evident the need of an accurate tool with a relatively low computational cost for the prediction of this kind of reacting flows. In the present work, a comparative study is carried out among large eddy simulations, performed with various grid resolutions, a Reynolds averaged Navier-Stokes simulation, and experimental data concerning the well-known Sandia D flame test case. In all the simulations, a flamelet progress variable model has been employed using various hypotheses for the joint probability density function closure. The filtered approach proved to be more accurate than the averaged one, even for the coarser grid used in this work. In fact both approaches have shown poorly accurate predictions in the first part of the combustion chamber, but only by the large eddy simulation one is capable to recover the inlet discrepancies with respect to the experimental data going along the streamwise direction.

*Keywords:* Large Eddy Simulation; Flamelet Progress Variable; Partialy premixed flame; Methane-Air combustion.

**Nomenclature**

| | |
|---|---|
| T | fluid temperature |
| U | fluid axial velocity |
| Z | mixture fraction |
| $\alpha$ | fluid molecular diffusivity |

* Corresponding author. Tel.: +390805963226; fax: +390805963411.
 *E-mail address:* pietro.depalma@poliba.it

| | |
|---|---|
| $\rho$ | fluid density |
| $\chi_i$ | mole fraction of the i[th] species |
| $\chi_z$ | scalar dissipation rate of the mixture faction |
| $\chi_{st}$ | scalar dissipation rate of the mixture faction computed at the stoichiometric point |
| $\omega_\phi$ | chemical production of the quantity $\phi$ |
| $\sim$ | Favre averaged variable |
| $\langle\ \rangle$ | Reynolds averaged variable |
| $'$ | Reynolds averaged variable fluctuation |
| $''^2$ | Variance |

## 1. Introduction

Even considering the large effort spent in the study of renewable energy resources, most of the energy used in industrial, civil and military applications comes from the combustion of hydrocarbons. Since, looking at the recent U.S. Energy Information Administration outlooks (2015), the importance of these fuels is going to rise over the next years, it appears worth developing tools able to improve the efficiency of the combustion processes.

In particular, modeling of turbulent combustion is one of the open problems in the field of reactive flow simulations. A direct approach, where the entire combustion phenomena is computed at runtime, is in fact inapplicable to flows at high Reynolds number such as those of practical interest. The possible choices developed nowadays for the solution of this problem can be categorized in two groups: those that aim at reducing the number of species in the system and those that reduce the kinetic mechanism into a functional manifold.

In particular, in the steady flamelet approach [1], which belongs to the second group, it is assumed that the Karlovitz number (the ratio between the chemical time scale and the Kolmogorov time scale) is much smaller than one, thus allowing the chemical proprieties to be computed in a space with a direction normal to the flame front and two directions tangential to it. For a diffusive flame, the advantage of using this frame of reference comes from expressing each thermo-chemical quantity as $\phi = \phi(Z, \chi_z)$, where $\chi_z = 2\alpha|\nabla Z|^2$. The functional form for $\phi$ used in this work is the solution of the steady flamelet equation:

$$-\rho \frac{\chi_z}{2} \frac{\partial^2 \phi}{\partial Z^2} = \omega_\phi. \tag{1}$$

In this way, for each stoichiometric scalar diffusion rate, it is possible to identify a solution composed of the distributions of each $\phi$, which will later be referred as flamelet. Solving this equation it is possible to obtain the well-known S-curve, shown in Fig. 1, where for each $\chi_{st}$ between the ignition and extinction points, the equation has three solutions branches: the non-ignited branch (coinciding with the x axis in Fig. 1); the metastable branch and the stable branch.

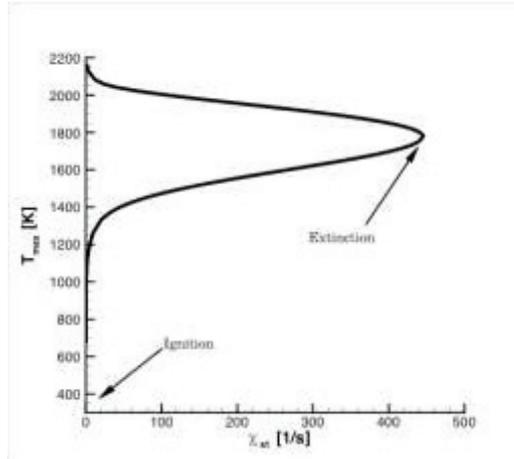

Fig. 1 S-curve for diluted methane-air combustion

An accurate approach to distinguish the three branches of the S-curve has been proposed by Pierce and Moin [2] and later developed by Ihme et. al. [3,4] and by Coclite et. al. [5,6] and is referred as the Flamelet Progress Variable (FPV) approach. The idea is to substitute $\chi_{st}$, in the mapping of $\phi$, with a progress variable $\Lambda$, that identifies the degree of completion of the combustion process. In this way it is possible to formulate the mapping of thermo-chemical quantities as $\phi = \phi(Z, \Lambda)$, where $\Lambda = C|Z_{flame}$ and $C$ is defined as the sum of the major combustion products mass fractions (in this case CO2, H2O, CO and H2). The procedure allows one to distinguish between the various solutions relative to the same stoichiometric dissipation rate. In addition to this model of the interaction between turbulence and chemistry, it has to be mentioned that, when the Navier-Stokes equations are solved using a turbulence model, namely when Reynolds Averaged Navier-Stokes (RANS) equations or Large Eddy Simulation (LES) are employed, the problem is formulated as function of the mass-weighted averages. Therefore, an additional model is required to evaluate the joint Favre averaged Probability Density Function (PDF) ($\tilde{P}(Z, C)$), which is needed to evaluate the mass-weighted average of $\phi$, as:

$$\tilde{\phi} = \iint \phi(Z, \Lambda)\tilde{P}(Z, \Lambda)dZd\Lambda \quad\quad\quad (2)$$

The shape of the joint PDF has to be presumed and the number of statistical moments of Z and C needed to determine the PDF are computed during the simulation by solving suitable additional transport equations.

The present work reports the validation of a newly developed LES code for the simulation of reactive flows using the FPV model by a comparison with the results obtained using an in-house recently developed RANS solver employing an original SMLD-FPV approach (see references [5,6,7] for the details of the RANS methodology). The flame Sandia D [8] is computed employing two grid densities and the LES code in order to provide a grid sensitivity study on this particular flame.

## 2 Mathematical Model

For the combustion test case considered in this work and the value of the Mach number well below the limit of 0.3, it has been possible to solve the Navier-Stokes equations decoupling the energy balance from the momentum and mass conservation equations. The density is therefore computed as a function of mixture composition and temperature. The continuity and momentum balance equations, filtered using a top-hat filter, are then solved using the second order accurate scheme of Ham et. al. [9] on an unstructured collocated grid. The sub-grid residual stress tensor is determined using the Germano et. al. [10,11] dynamic Smagorinsky model. At the inlet points, 500 Fourier modes have been superimposed to the mean velocity profile using the technique proposed by Keating and Piomelli [12] in order to match the experimental data provided by Schneider et. al. [13]. Although a sensitivity analysis of the flame structure to the inlet turbulent fluctuations is not available in literature, to the best of the authors' knowledge, matching the local Reynolds stress tensor is important to recover the correct mixing process inside the chamber.

As described in the introduction, the combustion phenomena will be modeled by the FPV approach. In particular the formulation used for the calculation of the joint PDF of Z and Λ is based on the hypothesis of statistical independence of the two scalars as:

$$\tilde{P}(Z,\Lambda) = \tilde{P}(Z)\tilde{P}(\Lambda|Z) = \beta(\tilde{Z},\widetilde{Z''^2})\delta(\tilde{\Lambda}), \qquad (3)$$

where the marginal PDF of Z and the conditional PDF of Λ have been assumed equal to a beta distribution and a delta function, respectively.

Thus, three additional scalar transport equations have been solved for $\tilde{Z}, \widetilde{Z''^2}$ and $\tilde{\Lambda}$ [4]. The turbulent diffusivity of the transported scalars has been calculated using the dynamic procedure proposed by Pierce and Moin [2]. In order to limit the computational cost of the evaluation of the residual diffusive stresses, they have been computed only for the $\tilde{Z}$ field and assumed equal for the other two scalars.

**3 Test case description**

The test case considered in the present work is the well-known Sandia D flame [8]. It consists of an open flame at atmospheric pressure where a mixture of methane and air (25-75%), injected from a nozzle of 7.2mm of diameter (later referred as $D_{nozzle}$) at a velocity of 49.6 m/s, reacts with a pilot composed of the products of a lean pre-combustion of methane with air. The pilot flow is injected through a nozzle with an outer diameter of 18.2mm at a mean flow velocity of 11.4 m/s. The pilot is then surrounded by a co-flow at about 1 m/s, as shown in Fig. 2. The Reynolds number based on the fuel flow properties and fuel nozzle diameter is equal to 22400.

The computational domain dimensions used in the simulation are: 80 $D_{nozzle}$, 27.5 $D_{nozzle}$ and $2\pi$, along the axial, radial and tangential directions, respectively. The numbers of points used for the two grids employed are summarized in Table 1. The radial number of points does not follow the same increment of the other directions because of the mesh topology used in the simulation. In fact, in order to avoid a particular centerline treatment, the central region of the control volume has been discretized using a Cartesian topology, whereas the rest is discretized as an "O" grid. For this reason, given a target on the mesh element size, the increment of points in the tangential direction allows one to use fewer points in the radial direction especially in the fuel nozzle region.

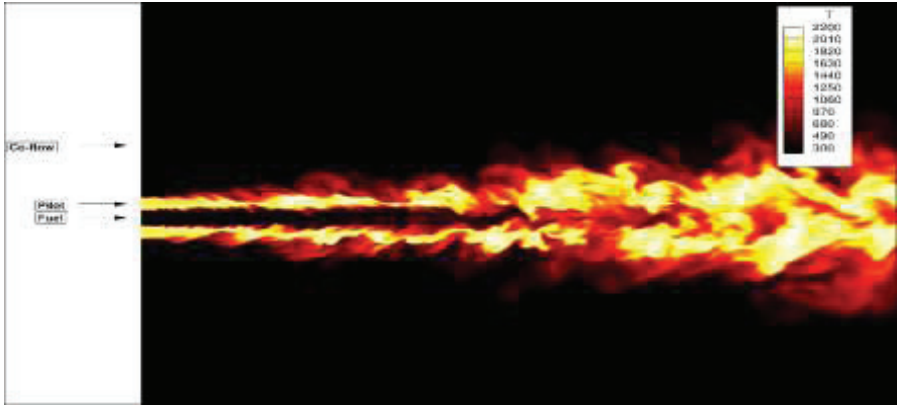

Fig. 2 LES using the finer grid: instantaneous temperature contour plot

Table 1. Summary of mesh size

| Grid name | Axial points | Radial points | Tangential points | Total cell. number |
|---|---|---|---|---|
| Coarse | 128 | 102 | 64 | 830000 |
| Fine | 340 | 130 | 64 | 2800000 |

All the simulations, initialized with velocity and scalar field equal to zero, have been run over 10 flow-through times in order to achieve the statistical stationary condition and then the time-averaged quantities have been computed over other 10 flow-through times.

**4. Results**

In this section, the results of the LES simulations will be compared with the results of the RANS simulations of Coclite et al. [6] and with experimental data [8,13] for validation. It is noteworthy that RANS simulations have been performed using a Statistically Most Likely Distribution (SMLD) approach to determine the joint Favre average PDF of Z and Λ. The effects of this model are still unknown in the LES and therefore this may constitute a source of discrepancy between the two sets of results.

*4.1 Profiles along the centerline*

Fig. 3 provides the Reynolds averages and Root Mean Squared (RMS) of temperature, mixture fraction and axial velocity component along the centerline up to 80 $D_{nozzle}$. There is a substantial agreement concerning the mean profiles between the simulations and the experimental data. LES appears to be more accurate, especially in the rear part of the combustion chamber. In particular considering the velocity profile, the Reynolds averaged simulations predict a steeper but lower decrease of the velocity magnitude in the middle of the combustion chamber with respect to the experimental data and to the LES.

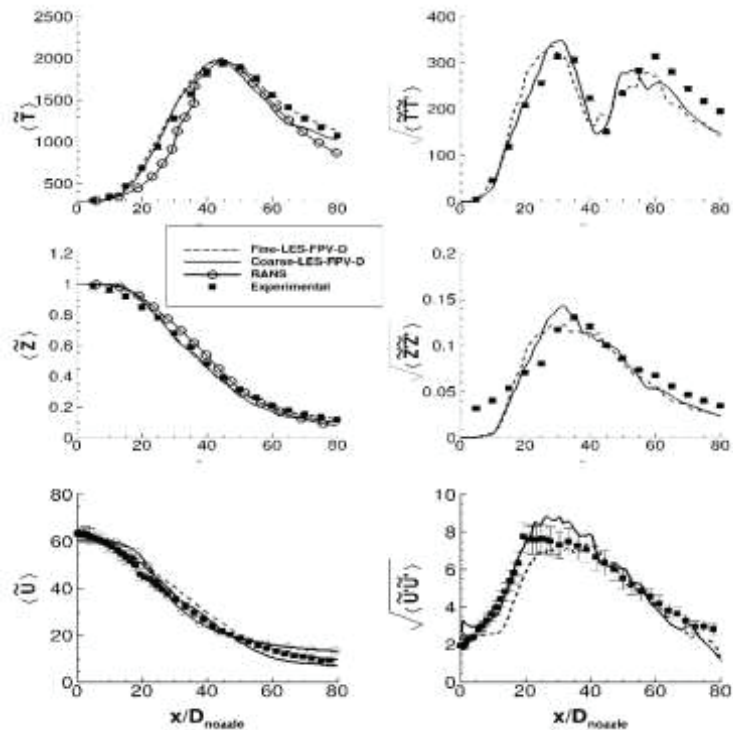

Fig. 3 Mean and RMS plot of temperature, mixture fraction and axial velocity along the centerline

The RMS are plotted only for the LES cases because they are not available for the RANS simulation. The largest discrepancy between the experimental and numerical data is obtained for the RMS of Z at the inlet section, due to the inlet condition imposed during the simulation. Moreover, further analyses are being performed in order to better understand the discrepancies observed in the RMS profiles of T at the chamber outlet and of U at 30 diameters from the inlet.

*4.2 Radial profiles*

Fig. 4 provides the radial profiles for Reynolds averaged temperature, mixture fraction and axial velocity at three different axial positions. Starting with the analysis of the temperature plots, it is evident that the LES slightly overestimates the energy released by combustion process, whereas the RANS underestimates it. In the first case, this effect disappears going through the combustion chamber, in fact in the last section the difference between the results and the experimental data is negligible. Instead, in the RANS simulation, this effect appears for all the analyzed sections. This result can be explained considering that the combustion process is mainly governed by diffusion and both approaches model the great part of the turbulent diffusion process. This also explains the reason why the LES accuracy improves along the combustion chamber. In fact, the turbulent length scales decrease with the distance from the fuel nozzle, letting the LES model to solve a larger part of the turbulent scales spectrum and to provide a better estimation of the diffusion processes. For the same reason, the field of the passive scalar Z is more accurate in the LES simulations.

Furthermore, in these plots it is possible to note that the LES provides a velocity field that is closer to the experimental data with respect to the RANS.

## 5. Conclusions

In the present work a comparison among reactive flow simulations performed using different turbulence modeling approaches has been provided in order to assess the effective gain in accuracy of LES with respect to RANS simulations for partially premixed methane-air combustion. The LES simulation has been coupled with a standard FPV combustion model, whereas the RANS employs a more accurate SMLD approach.

The LES has a significant increase of computational cost, even using a relatively coarse grid. However, it has been shown that, at least for the considered test case, the increased accuracy in the LES flow description overcomes the performance of the RANS-SMLD model.

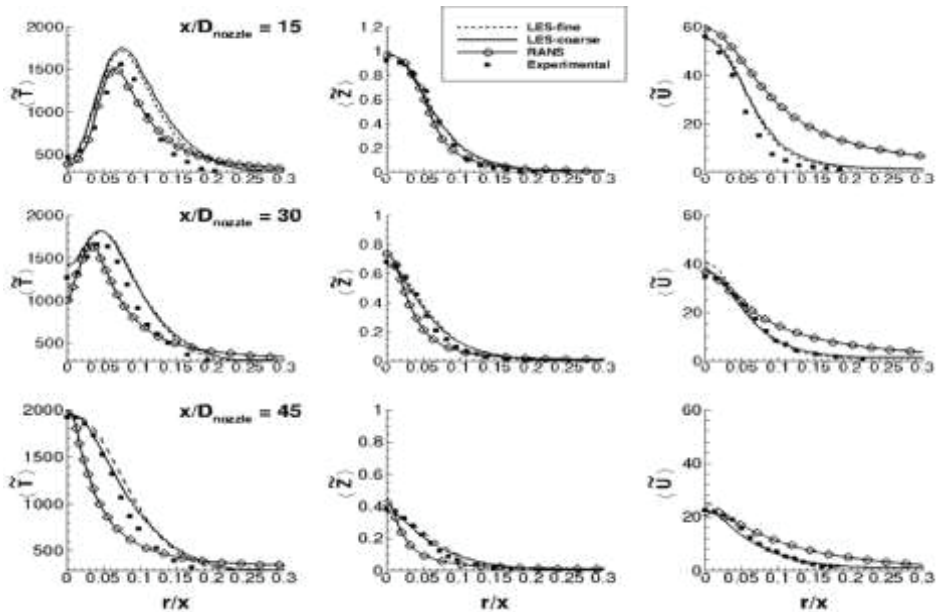

Fig. 4 Radial profiles of mean and rms plot of temperature, mixture fraction and axial velocity at $x/D_{nozzle}$=15,30,45

It is still a trade-off problem, between computational time and flow description accuracy, to decide whether it is better to use an LES or RANS for practical cases. However, looking at the differences observed in this work, it seems that even a low-resolved LES can be more accurate than a RANS simulation for this kind of flows, with an acceptable computational cost.

Future work will aim at extending the recently developed joint-PDF SMLD approach, already employed in conjunction with the RANS model, to the large eddy simulation.

**Acknowledgements**

This research has been supported by grant n. *PON03PE* 00067 6 APULIA SPACE.

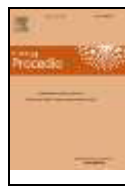

**Biography**

Pietro De Palma was born in Bari (Italy) in 1966. He obtained the Diploma Course at the VKI in 1990 and the Ph. D. in Mechanical Engineering at Politecnico di Bari where, from December 2003, he is professor of hydraulic and thermal machines.